\newcommand{\be}{\begin{equation}}\newcommand{\ee}{\end{equation}}
\newcommand{\bea}{\begin{eqnarray}}\newcommand{\eea}{\end{eqnarray}}
\newcommand{\nn}{\nonumber}\newcommand{\p}[1]{(\ref{#1})}
\newcommand{\lb}[1]{\label{#1}}
\newcommand\s{\scriptscriptstyle}

\newcommand\q{\quad}
\newcommand\qq{\quad\quad}
\renewcommand\={\ =\ }

\newcommand\cC{{\cal C}}

\newcommand\cN{{\cal N}}

\newcommand\cW{{\cal W}}
\newcommand\bcW{\bar{\cal W}}

\newcommand\stc{\stackrel{\star}{,}}

\newcommand\olp{\overleftarrow{\partial}}
\newcommand\orp{\overrightarrow{\partial}}
\newcommand\olD{\overleftarrow{D}}
\newcommand\orD{\overrightarrow{D}}
\newcommand\olbD{\overleftarrow{\bar D}}
\newcommand\orbD{\overrightarrow{\bar D}}

\newcommand\tpa{\theta^{+\alpha}}

\newcommand\tma{\theta^{-\alpha}}

\newcommand\tka{\theta^{\alpha}_k}

\newcommand\tjb{\theta^{\beta}_j}
\newcommand\tib{\theta^{\beta}_i}

\newcommand\btka{\bar\theta^{\da k}}
\newcommand\btkb{\bar\theta^{\db k}}

\newcommand\btjb{\bar\theta^{\db j}}
\newcommand\btja{\bar\theta^{\da j}}

\newcommand\btpa{\bar{\theta}^{+\dot{\alpha}}}
\newcommand\btpb{\bar{\theta}^{+\dot{\beta}}}

\newcommand\btma{\bar{\theta}^{-\dot{\alpha}}}

\newcommand\tp{\theta^+}

\newcommand\btp{\bar\theta^+}

\newcommand\ada{{\alpha\dot{\alpha}}}
\newcommand\adb{{\alpha\dot{\beta}}}
\newcommand\bdb{{\beta\dot{\beta}}}

\newcommand\ab{{\alpha\beta}}
\newcommand\ba{{\beta\alpha}}

\newcommand\da{{\dot{\alpha}}}
\newcommand\db{{\dot{\beta}}}
\newcommand\dr{{\dot{\rho}}}

\newcommand\A{{\s A}}

\newcommand\R{{\s R}}
\newcommand\M{{\s M}}
\newcommand\sL{{\s L}}

\newcommand\W{{\s W}}
\newcommand\Z{{\s Z}}

\newcommand{\pp}{{\s ++}}
\newcommand{\m}{{\s --}}

\newcommand{\Dp}{D^{\pp}}
\newcommand{\Dm}{D^{\m}}

\newcommand{\dpp}{\partial^{\pp}}

\newcommand{\Vp}{V^\pp}
\newcommand{\Vm}{V^\m}

\newcommand{\Dpa}{D^+_\alpha}
\newcommand{\Dma}{D^-_\alpha}
\newcommand{\Dpb}{D^+_\beta}

\newcommand{\bDpa}{\bar{D}^+_{\dot{\alpha}}}
\newcommand{\bDpb}{\bar{D}^+_{\dot{\beta}}}
\newcommand{\bDma}{\bar{D}^-_{\dot{\alpha}}}

\def\sfrac#1#2{{\textstyle\frac{#1}{#2}}}
\def\ha{\frac12}
\def\sha{\sfrac12}

\def\e{\mbox{e}}
\def\ii{\mbox{i}}
\def\diff{\mbox{d}}

\documentclass[12pt]{article}
\usepackage{amsmath,amssymb}

\begin{document}

\thispagestyle{empty}

\begin{center}
{\Large\bf
Nilpotent Deformations of N=2 Superspace}
\vspace{2cm}

{\large\bf
Evgeny Ivanov${\ }^{a}$, \
Olaf Lechtenfeld${\ }^{b}$, \
Boris Zupnik${\ }^{a}$ }
\end{center}

\vspace{0.2cm}

\begin{center}
${}^a$ {\it Bogoliubov Laboratory of Theoretical Physics, 
JINR, 141980 Dubna, Russia}\\
{\tt eivanov@thsun1.jinr.ru, zupnik@thsun1.jinr.ru}
\end{center}

\begin{center}
${}^b$ {\it Institut f\"ur Theoretische Physik, Universit\"at Hannover,
30167 Hannover, Germany} \\
{\tt lechtenf@itp.uni-hannover.de}
\end{center}

\vspace{2cm}

\begin{abstract}
\noindent
We investigate deformations of four-dimensional $N{=}(1,1)$
euclidean superspace induced by nonanticommuting fermionic coordinates.
We essentially use the harmonic superspace approach and consider nilpotent
bi-differential Poisson operators only. One variant of such deformations
(termed chiral nilpotent) directly generalizes the recently studied
chiral deformation of $N{=}(\ha,\ha)$ superspace. It preserves chirality
and harmonic analyticity but generically breaks $N{=}(1,1)$ to $N{=}(1,0)$
supersymmetry. Yet, for degenerate choices of the constant deformation matrix
$N{=}(1,\ha)$ supersymmetry can be retained, i.e.~a fraction of~3/4.
An alternative version (termed analytic nilpotent) imposes minimal
nonanticommutativity on the analytic coordinates of harmonic superspace.
It does not affect the analytic subspace and respects all supersymmetries,
at the expense of chirality however. For a chiral nilpotent deformation,
we present non(anti)commutative euclidean analogs of $N{=}2$ Maxwell and
hypermultiplet off-shell actions.

\end{abstract}
\newpage
\hrule
{\bf Contents}\\

{\bf 1. Introduction\hfill 1}\\

{\bf 2. Deformation of $N=2$ chiral superspace\hfill 3}\\

{\bf 3. Deformations of $N=2$ harmonic superspace\hfill 8}\\

{\bf 4. Interaction in deformed harmonic superspace\hfill 11}\\

{\bf 5. Conclusions\hfill 13}\\

\hrule

\setcounter{page}{1}
\section{Introduction}

Moyal-type deformations of superfield theories are currently a subject of
intense study (see, e.g.~\cite{BrSc}--\cite{BS}).
Analogous to noncommutative field theories on bosonic spacetime,
noncommutative superfield theories can be formulated in ordinary superspace
by multiplying functions given on it via a star product which is generated
by some bi-differential operator or Poisson structure~$P$.
The latter tells us directly which symmetries of the undeformed (local)
field theory are explicitly broken in the deformed (nonlocal) case.

Generic Moyal-type deformations of a superspace are characterized by
a constant graded-antisymmetric non\-(anti)\-commutativity matrix
$(C^{AB})\,$.\footnote{
The interpretation of quantizing fermionic systems
as a deformation of the Grassmann algebra, in analogy with the Weyl-Moyal
quantization of bosonic systems, can be traced back to~\cite{Be,Ca}.}
A minimal deformation of euclidean $N{=}1$ superspace -- more suitably denoted
as $N{=}(\ha,\ha)$ superspace -- was considered in a recent paper~\cite{Se}.
For the chiral $N{=}1$ coordinates $(x^m_\sL,\theta^\alpha,\bar\theta^\da)$
the non\-(anti)\-commutativity was restricted to
\be
\{\theta^\alpha\stc \theta^\beta \}\ :=\
\theta^\alpha \star \theta^\beta + \theta^\beta \star \theta^\alpha \=C^\ab\ ,
\lb{n1}
\ee
i.e.~the basic star products read
\be
\theta^\alpha\star\theta^\beta\=\theta^\alpha\,\theta^\beta+\sha C^\ab\ ,\quad
\theta^\alpha\star x^m_\sL \= \theta^\alpha\,x^m_\sL\ ,\quad
x^m_\sL\star x^n_\sL \= x^m_\sL\,x^n_\sL\ ,
\lb{n2}
\ee
with $(C^\ab)$ being some constant symmetric matrix.
Note that the bosonic and the antichiral coordinates have undeformed
commutation relations with everyone, so $(C^{AB})$ is rather degenerate here.
For functions $A$ and $B$ of $(x^m_\sL,\theta^\alpha,\bar\theta^\da)$
the star product~\p{n2} is generated as
\be
A \star B \= A\,\e^P B \= A\,B + A\,P\,B + \sha A\,P^2 B
\ee
where the bi-differential operator
\be
P\=-\sha\olp_{\!\alpha} C^\ab \orp_{\!\beta}
\qq\textrm{is nilpotent:}\q P^3\=0\ .
\lb{P1}
\ee
This defines a particular case of a deformed superspace.
It retains $N{=}(\ha,0)$ of the original
$N{=}(\ha,\ha)$ supersymmetry because $Q_\alpha$ commutes with~$P$ while
$\bar{Q}_\da$ does not. Deformations generated by a nilpotent Poisson structure
such as~\p{P1} will be called {\it nilpotent deformations\/} in
this paper.

Some quantum calculations in this deformed superspace, in particular
concerning non-renormalization theorems, were presented
in~\cite{BFR}--\cite{GPR}. The string-theoretic origin of this type of
noncommutativity was discussed in~\cite{ova,Se,BeS}
(see also~\cite{BrSc,vN,HIU}).  More recently,
the issue of analogous deformations of $N{=}2$ extended superspace
in four dimensions was addressed in~\cite{BeS}.

While the preservation of chirality is the fundamental underlying principle
of $N{=}1$ superfield theories~\cite{F0}, it is the power of {\it Grassmann
harmonic analyticity\/} which replaces the use of chirality in $N{=}2$
supersymmetric theories in four dimensions~\cite{GIO}--\cite{GIOS}.
Therefore, it is natural to look for nilpotent deformations of $N{=}(1,1)$
Euclidean superspace which preserve this harmonic analyticity (perhaps in
parallel with chirality). The basic aim of the present paper is to describe
such deformations and to give deformed superfield actions for a few textbook
examples of $N{=}2$ theories. We analyze the role of the standard conjugation
or an alternative pseudoconjugation in euclidean $N{=}2$ supersymmetric
theories and their deformations.

Section 2 generalizes Seiberg's simplest nilpotent deformation to euclidean
$N{=}(1,1)$ superspace.
We call this a {\it chiral nilpotent deformation\/}.
The corresponding bi-differential operator~$P$ acts on the left half
of the $N{=}(1,1)$ spinor coordinates only and preserves all chiralities.
It retains $N{=}(1,0)$ supersymmetry but generically breaks the R-symmetry
SU(2) as well as the Euclidean invariance SO(4)$\,\to\,$SU$(2)_R$.
For judicious choices, however, $N{=}(1,\ha)$ supersymmetry or the whole
automorphism group SO(4)$\times\,$SU(2) can be preserved.

In Section 3 we reformulate the chiral nilpotent deformation in the euclidean
version of $N{=}2$ harmonic superspace~\cite{GIK1}--\cite{GIOS} and show that
it preserves Grassmann harmonic analyticity. In the analytic coordinates, the
chiral deformation acts not only on the left-handed fermionic but also on the
bosonic coordinates (albeit still in nilpotent fashion). This remains true
when we restrict to the analytic subspace, i.e.~consider Grassmann analytic
superfields. As an interesting alternative, there exist nilpotent deformations
which affect neither anti-chiral superfields nor analytic superfields but act
in the central superspace. We call them {\it analytic nilpotent deformations\/}.
In this case the analytic subspace is undeformed but chirality is no longer
preserved, still leading to deformed products for general superfields e.g.~in
$N{=}(1,1)$ super Yang-Mills. Remarkably, the full $N{=}(1,1)$ supersymmetry
remains intact here. This option does not exist in $N{=}1$ superspace.

Noncommutative interactions of harmonic superfields are considered in Section~4.
Formally these interactions resemble those in the purely bose-deformed harmonic
superspace of~\cite{BS}. As an important difference though, the nilpotency of
our deformations is expected to render quantum calculations much more feasible.

The main novel developments in our work are the analysis of euclidean $N{=}2$
supersymmetry breaking deformations in harmonic superspace and the construction
of the relevant superfield models.  The supersymmetry-preserving
deformations of $N{=}2$ superspace were considered in \cite{FL,KPT,BS}. While
writing this paper, a preprint \cite{FLM} appeared which discusses
supersymmetry-breaking deformations of $N{=}2$ superspace on equal footing
with supersymmetry-preserving ones.

\setcounter{equation}0
\section{Deformations of N=2 chiral superspace}

In the deformation~\p{n1} of $N{=}1, D{=}4$ euclidean superspace proposed
in~\cite{Se} one introduces non\-(anti)\-commutativity only for one half
of the spinor coordinates. By construction, this deformation preserves the
chiral representations of $N{=}1$ supersymmetry.
The bi-differential operator of~\cite{Se} has the form~\p{P1}
and acts on standard superfields $V(x^m_\sL,\theta^\alpha,\bar\theta^\da)$.

It is important to realize that the (pseudo)conjugation properties of spinors
in 4D Euclidean space with the group $Spin(4)$=SU$(2)_\sL\times\,$SU$(2)_\R$
are radically different from those in Minkowski space since left- and
right-handed SU(2) spinors are independent. A conjugation or pseudoconjugation
is defined as an antilinear map $\sigma$ which acts in the algebra of complex
superfields via $\sigma(AB)=\sigma(B)\,\sigma(A)$ and is reduced to the standar
complex conjugation on complex numbers. Furthermore, a conjugation satisfies
$\sigma\bigl(\sigma(A)\bigr)=A$ while a pseudoconjugation obeys
$\sigma\bigl(\sigma(A)\bigr)=(-1)^{|A|}A$ where $|A|$ is some $Z_2$-grading of
$A$. For $N{=}1$  euclidean superspace to have the real dimension $(4|4)$
like its Minkowski counterpart, one is led to apply the following
pseudoconjugation of the  SU$(2)_\sL\times\,$SU$(2)_\R$
spinor Grassmann coordinates (see e.g. \cite{Ma}):~\footnote{
We use the conventions $\varepsilon_{12}=-\varepsilon^{12}=
\varepsilon_{\dot{1}\dot{2}}=-\varepsilon^{\dot{1}\dot{2}}=1$ and
$\sigma_m=(I,\ii\overrightarrow\sigma)$ for the basic quantities in
euclidean space.
}
\be
(\theta^\alpha)^*\=\varepsilon_\ab\theta^\beta\ ,\qq
(\bar\theta^\da)^*=\varepsilon_{\da\db}\bar\theta^\db\,. \label{Oo}
\ee
Here, the map  ${}^*$ is a pseudoconjugation which squares to $-1$ on any
odd $\theta$ monomial (and on the fermionic component fields) and to $+1$
on any even monomial (and on the bosonic component fields). So, when acting
on bosonic fields, it is indistinguishable from the standard complex
conjugation. It preserves the irreducible representations
of the group  $Spin(4)$ (it is straightforward to check that \p{Oo}
is consistent with the action of this group). Examples of bilinear
real combinations of $N{=}1$ euclidean spinors are
\be
\ii\,\theta\,\psi(x)\quad,\qquad \ii\,\theta\sigma^m\bar\theta\quad,\qquad
\ii\,\psi(x)\sigma_m\partial_m\bar\psi(x)
\ee
where $\psi^\alpha(x)$ and $\bar\psi^\da(x)$ are pseudoreal spinor fields (they
satisfy the conditions \p{Oo}). As an important
consequence of the pseudoreality of spinor  coordinates, $N{=}1$ euclidean
chiral superfields can be chosen as real (like the general ones).

Though our main goal is to introduce consistent {\it nilpotent\/} deformations
of $N{=}(1,1)$ {\it harmonic\/} superspace, it is convenient to start the analysis
in ordinary $N{=}(1,1)$ superspace in {\it chiral\/} parametrization.
We consider euclidean $N{=}(1,1)$ superspace and use the chiral coordinates
\be
z_\sL\ \equiv\ ( x^m_\sL, \tka, \btka )
\ee
which transform under $N{=}(1,1)$ supersymmetry as
\be
\delta_\epsilon  x^m_\sL\=2\ii(\sigma^m)_\ada\tka\bar\epsilon^{\da k}\ ,\qq
\delta_\epsilon \tka\=\epsilon^\alpha_k\ ,\qq
\delta_\epsilon \btka\=\bar\epsilon^{\da k}\ ,
\lb{standSUSY}
\ee
where $\epsilon^\alpha_k$ and $\bar\epsilon^{\da k}$ are the transformation
parameters.
The `central' bosonic coordinate $x^m$ is related to the `left' coordinate by
\be
x^m_\sL\=x^m+\ii(\sigma^m)_\ada\tka\btka\ .
\lb{xLx}
\ee
As automorphisms we have the euclidean space spinor group
$Spin(4)$ and the R-symmetry group
SU(2)$\times\,$O(1,1)  acting simultaneously on
left and right spinors.

Let us dwell in some detail on the (pseudo)reality properties of
$N{=}(1,1)$ superspace. We assume the Grassmann coordinates to be real
with respect to the standard conjugation
 \be
 \widetilde{\tka}\=\varepsilon^{kj}\varepsilon_\ab\theta_j^\beta\ ,\qq
\widetilde{\btka}\=-\varepsilon_{kj}\varepsilon_{\da\db}\bar\theta^{\db j}
\ ,\qq
\widetilde{x^m_\sL}\={x^m_\sL}\ ,\qq
\widetilde{AB}\=\widetilde{B}\widetilde{A}\ .
\lb{ELconj}
\ee
This conjugation squares to the identity on any object, and with respect to 
it
the  $N{=}(1,1)$ superspace has the real dimension $(4|8)$.
The component spinor fields enjoy the analogous conjugation properties.
Eq.~(\ref{ELconj}) is evidently compatible with both $Spin(4)$ and
R-symmetries, preserving
any irreducible representation of these groups. However, the $N{=}(\ha,\ha)$
superspace cannot be treated as a real subspace of the $N{=}(1,1)$ superspace
if one considers only this standard conjugation.

Surprisingly, in the same  euclidean $N{=}(1,1)$ superspace one can define
an analog of the pseudoconjugation \p{Oo}, namely
\be
(\tka)^*\=\varepsilon_\ab\theta^\beta_k\ ,\qq
(\btka)^*\=\varepsilon_{\da\db}\bar\theta^{\db k}\ ,\qq
(x^m_\sL)^*\=x^m_\sL\ ,\qq
(AB)^*\=B^* A^*
\lb{sr2conj}
\ee
with respect to which the $N{=}(\ha,\ha)$ superspace forms a real subspace.
The existence of this pseudoconjugation does not imply any further restriction
on the $N{=}(1,1)$ superspace.
It preserves representations of $N{=}(1,1)$ supersymmetry
and, like \p{Oo}, is compatible with the action of the group $Spin(4)$.
It is also compatible with the R-symmetry group O(1,1). As for the R-symmetry
group SU(2), it preserves only some U(1) subgroup of the latter.
In other words, the standard conjugation \p{ELconj} and the
pseudoconjugation \p{sr2conj} act differently on objects transforming under
non-trivial representations of this SU(2), e.g. on Grassmann coordinates.
The map ${}^*$ squares to $-1$ on these coordinates and the associated
spinor fields, and to $+1$ on any bosonic monomial or field.
Yet, even on bosonic objects
the two maps act in a different way if these objects belong to a non-trivial
representation of the R-symmetry SU(2) (see eq. \p{Ooo} below). Only on
the singlets of the latter, e.g. scalar $N{=}(1,1)$ superfields and R-invariant
differential operators, both maps act as the standard complex conjugation.
In particular, the invariant actions are real with respect to both
${}^*$ and ${}^\sim$, despite the fact that the component fields may
have different properties under these
(pseudo)conjugations. Clearly, it is the pseudoconjugation ${}^*$ which is
respected by the reduction $N{=}(1,1) \;\rightarrow\; N{=}(\ha,\ha)$. Such a
reduction preserves the pseudoreality but explicitly breaks the SU(2) R-symmetry.

In chiral coordinates, a {\it chiral nilpotent deformation\/} for products
of superfields is determined by the following operator,
\bea
P &=&-\sha\olp{}^k_\alpha\, C^\ab_{kj} \orp{}^j_\beta
\=-\sha\overleftarrow{Q}{}^k_\alpha\,C^\ab_{kj}\overrightarrow{Q}{}^j_\beta
\qq\textrm{such that} \nn\\[8pt]
A P B &=& -\sha(A\olp{}^k_\alpha) C^\ab_{kj} (\orp{}^j_\beta B)
\=-\sha(-1)^{p(A)}(\partial^k_\alpha A)C^\ab_{kj}(\partial^j_\beta B)\nn\\[6pt]
&=& -(-1)^{p(A)p(B)}B P A\ .
\lb{Poper}
\eea
Here, $C^\ab_{kj}=C^\ba_{jk}$ are some constants,
$\;p(A)$ is the supersymmetry $Z_2$-grading,
while $Q^k_\alpha=\partial^k_\alpha$ are the generators of left supersymmetry
and the derivatives act as
\be
\partial^k_\alpha\tib=\delta_i^k\delta^\beta_\alpha
\qq\textrm{and}\qq
\bar\partial_{\da i}\btkb=\delta^k_i\delta^\db_\da\ .
\ee
By definition, the operator \p{Poper} preserves both chirality and
anti-chirality and does not touch the SU$(2)_R$.
It induces a graded Poisson bracket on superfields \cite{FL,KPT}.
We also demand $P$ to be real, i.e.~invariant under some antilinear map
in the algebra of superfields.
The two possible (pseudo)conjugations introduced above then lead to
different conditions
\bea
&& \p{sr2conj} \qq\Longrightarrow\qq (C^\ab_{kj})^*\=C_{\ab kj} \\
&& \p{ELconj} \qq\Longrightarrow\qq \widetilde{C^\ab_{kj}}\=C_\ab^{kj}\ .
\eea
Since $(APB)^*=B^*PA^*$ and $\widetilde{A P B}=\widetilde{B}P\widetilde{A}$,
our star-product satisfies the following natural  rules:
\be
(A \star B)^* \= B^* \star A^*\ ,\qq
\widetilde{(A \star B)} \= \widetilde B \star \widetilde A\ .
\lb{Pconj}
\ee
Under SU$(2)_L\times\,$SU(2), the constant deformation matrix~C decomposes into
a (3,3) and a (1,1) part (see also~\cite{FLM}),
\be
C^\ab_{kj}\=C^{(\ab)}_{(kj)}+\varepsilon^\ab\varepsilon_{kj}I\ .
\lb{const}
\ee

It is worth pointing out that the (1,1) part preserves the full
SO$(4)\times\,$SU(2) symmetry. The (3,3) part may be diagonalized by employing
SU$(2)_L\times\,$SU(2) rotations, so that it can be brought to a minimal form of
\be
C^{(\ab)}_{(kj)} \= \delta^{(\ab)}_{(kj)}\,C^{(\ab)} \ .
\ee
Depending on the signs (or vanishing) of the constants $C^{(\ab)}$,
after rescaling of $\tka$ this yields the Clifford algebra in three
or in lower dimensions. In an operator realization, therefore,
$\tka$ are proportional to Pauli matrices or fermionic creation/annihilation
operators. This complements the bosonic oscillator representation of
(non-nilpotently) noncommuting bosonic coordinates in their Darboux basis.

Note that the manifestly $N{=}2$ supersymmetric bi-differential operators
of~\cite{FL,KPT} involve flat spinor derivatives $D^k_\alpha$ instead of
partial derivatives. This choice violates chirality. In contrast,
we basically follow the line of~\cite{Se} and
investigate deformations which preserve irreducible representations
based on chirality and/or Grassmann harmonic analyticity (see Section~3),
but may explicitly break some fraction of supersymmetry.

Given the operator \p{Poper}, the Moyal product of two superfields reads
\bea
A \star B &=& A\,\e^P B \=
A\,B+ A\,P\,B + \sha A\,P^2 B + \sfrac16 A\,P^3 B + \sfrac{1}{24} A\,P^4 B \nn\\
&=& A\,B + \partial^k_\alpha N_k^\alpha(A,B,C)\ ,
\lb{moyal}
\eea
where the identity $P^5=0$ was used and $N_k^\alpha(A,B,C)$ is some function
of the superfields and the constants $C^\ab_{kj}$. By construction, this
product is associative and satisfies the standard Leibniz rule
\bea
&&\partial_\M (A \star B)\=
\partial_\M A \star B + (-1)^{p(M)p(A)}A \star \partial_\M B
\eea
for partial derivatives
$\partial_\M =(\partial_m^\sL, \partial^k_\alpha, \bar\partial_{\da k})\lb{pL}$
and equally for any differential operator not containing $\tka$. In particular,
this star product preserves antichirality, $D^k_\alpha\bar\Phi=0$.
{}From~\p{moyal} it is easy to see that the chiral-superspace integral
of the Moyal product of two superfields is not deformed,
\bea
\int\!\diff^4x\, \diff^4\theta\; A \star B \=
\int\!\diff^4x\, \diff^4\theta\; A\, B\ ,
\eea
while integrals of star products of three or more superfields are deformed.
For the superspace coordinates we obtain the following graded star-product
commutators,
\bea
&&\{\tka \stc \tjb\} \=  C^\ab_{kj}\ ,\qq
[x^m_\sL \stc x^n_\sL] \=0\ ,\qq
[x^m_\sL \stc \tka] \=0\ ,\nonumber\\[6pt]
&&[x^m_\sL \stc \btka] =0\, \qq
\{\tka \stc \btjb\} \=0\ ,\qq
\{\btka \stc \btjb\} \=0\ . \lb{basic}
\eea

The $N{=}(1,1)$ supersymmetry is realized in the standard way on all
superfields, i.e.~the supersymmetry transformations of the component fields
are undeformed. At the same time, the star product~\p{moyal} of two superfields
is not a fully covariant object because the operator $P$ of~\p{Poper} breaks
some of the symmetries.
Thus, in our treatment only free actions preserve all supersymmetries
while interactions get deformed and are not invariant under all standard
supersymmetry transformations.
To exhibit the residual symmetries of a deformed interacting theory,
we formulate the invariance condition
\be
[K, P]\=0
\ee
for the corresponding generators~$K$ in the standard $N{=}(1,1)$ superspace.
Clearly, this condition is generically not met by differential operators
depending on~$\tka$, such as
\bea
&&L^\alpha_\beta\=
\sha (\sigma_{mn})^\alpha_\beta x^m_\sL\partial^n_\sL +
\theta^\alpha_k \partial^k_\beta -
\sha\delta^\alpha_\beta\theta_k^\rho \partial^k_\rho\ ,\nn\\[6pt]
&&L^k_j\=
\theta^\alpha_j\partial^k_\alpha -
\sha\delta^k_j\theta_i^\rho\partial^i_\rho -
\bar\theta^{\da k}\bar\partial_{\da j} +
\sha\delta^k_j\bar\theta^{\dr i} \bar\partial_{\dr i}\ ,\nn\\[6pt]
&&\bar Q_{\da k}\=
\bar\partial_{\da k} - 2\ii(\sigma^m)_\ada\tka\partial^\sL_m\ ,
\lb{defQ}
\eea
and the symmetries generated by these are explicitly broken in the deformed
superspace integrals.
Out of all supersymmetry and automorphism generators, only $Q_\alpha^k$ and
$\bar L^\da_\db$ do commute with~$P$ of~\p{Poper}.
Hence, for a generic choice of the constant matrix $(C^{\ab}_{ik})$,
the breaking pattern is $N{=}(1,1)\to N{=}(1,0)$ for supersymmetry and
SO$(4)\times\,$O$(1,1)\times\,$SU$(2)\,\to\,$SU$(2)_R$
for Euclidean and R-symmetries.

An exception occurs for the singlet part in \p{const}, i.e. for
\be
C^{(\ab)}_{(kj)}\=0 \qq\Longrightarrow\qq
C^\ab_{kj}\=\varepsilon^\ab\varepsilon_{kj} I
\qq\Longleftrightarrow\qq
P_{singlet}\=-\sha\,\overleftarrow{Q^k_\alpha}\,I\,\overrightarrow{Q^\alpha_k}
\ , \lb{Cinv}
\ee
which is fully SO$(4)\times\,$SU(2) invariant and non-degenerate but also
fully breaks the right half of supersymmetry.

Is it possible to break less than one half of the supersymmetry?
The answer depends on our choice of conjugation.
The $\ \ \widetilde{}\ \ $ conjugation \p{ELconj} connects $Q^1_\alpha$
with $Q^{\alpha 2}$, so the condition $\widetilde{P}{=}P$
necessarily breaks $N{=}(0,1)$ supersymmetry.
The choice of the ${}^*$ pseudoconjugation \p{sr2conj}, on the other hand,
is compatible with the decomposition of $N{=}(1,1)$ into two
$N{=}(\ha,\ha)$ superalgebras, each given by a fixed value for the SU(2)~index.
Therefore, it allows one to pick a degenerate deformation, e.g.
\be
P_{deg}(Q^2)
\=-\sha C^{12}_{22} (\overleftarrow{Q}{}^2_1\,\overrightarrow{Q}{}^2_2
+ \overleftarrow{Q}{}^2_2\,\overrightarrow{Q}{}^2_1)\ ,\lb{1break}
\ee
which does not involve $Q^1_\alpha$.
In this case, only $\bar Q_{\da 2}$ are broken but not the supercharges
$\bar Q_{\da 1}$. Hence, the deformation $P_{deg}$
preserves $N{=}(1,\ha)$ supersymmetry.

It is of course possible to consider more general deformations affecting
both the chiral sector and the anti-chiral one.
Since in euclidean space left and right sectors do not talk to each other,
we may combine a chiral deformation~$P$ with some anti-chiral one, $\bar{P}$.
The only (though rather restrictive) requirement is that $\bar{P}$ commutes
with~$P$. Otherwise, the important associativity property for the resulting
deformation would be ruined.
For instance, one may add to $P$ of~\p{Poper} any one of the following
bi-differential operators,
\be
L\= -\sha \olD{}^k_\alpha\,J\,\orD{}_k^\alpha \qq \mbox{or} \qq
R\= -\sha \olbD{}_{\da k}\,\bar{J}\,\orbD{}^{\da k}
 \= -\sha \olp{}_{\da k}\,\bar{J}\,\orp{}^{\da k}\ ,\lb{LRdef}
\ee
where $J$ and $\bar J$ are some real constants.
Each of these commutes with only one of the two chirality operators,
but preserves the full supersymmetry and yields an SO$(4)\times\,$SU(2)
invariant deformation.
On their own, these operators produce deformations for general superfields
without any symmetry breaking and so belong to the class of fully
supersymmetric Moyal operators considered in \cite{FL,KPT,FLM}.
The deformation operator $P{+}R$ or $P{+}L$ breaks the same portion of
(super)symmetry as $P$ does. It preserves chirality or anti-chirality,
respectively. Since $L$ and $R$ induce graded star-commutators more general
than the minimal set~\p{basic}, the corresponding deformations lie outside
the class of chiral nilpotent ones.
We shall encounter them again in the following section.

\setcounter{equation}0
\section{Deformations of N=2 harmonic superspace}

The basic concepts of the $N{=}2, D{=}4$ harmonic superspace \cite{GIK1,GIOS2}
are collected in the book \cite{GIOS}. The spinor $SU(2)/U(1)$ harmonics
$u^\pm_i$ subject to
\be
u^{\pm k}\=\varepsilon^{kj}u^\pm_j
\qq\textrm{and}\qq
u^{+k} u^-_k\= 1
\qq\textrm{with}\qq
\delta_\epsilon u^{\pm k}\= 0
\ee
can be used to construct analytic coordinates
$(x_\A^m,\theta^{\pm\alpha},\bar\theta^{\pm\da})$
in the euclidean version of $N{=}2$ harmonic superspace:
\bea
&& x^m_\A\=
x^m_\sL-2\ii(\sigma^m)_\ada\theta^{\alpha k}\btja u^{-}_ku^+_j\ ,\qq
\delta_\epsilon x^m_\A\= -2\ii(\sigma^m)_\ada
(\epsilon^{-\alpha} \btpa + \tpa\bar\epsilon^{-\da})\ , \nn\\[6pt]
&&\theta^{\alpha\pm}\= \theta^{\alpha k}u^\pm_k\ ,\ \q
\bar\theta^{\pm\da} \= \bar\theta^{\da k}u^\pm_k\ ,\qq
\delta_\epsilon \theta^{\alpha\pm}\= \epsilon^{\pm\alpha}\ ,\ \q
\delta_\epsilon\bar\theta^{\pm\da}\= \bar\epsilon^{\pm\da} \ ,
\lb{ancoor}
\eea
where $\epsilon^{\pm\alpha}{=}\epsilon^{\alpha k}u^\pm_k~,~
\bar\epsilon^{\pm\da}{=}\bar\epsilon^{\da k}u^\pm_k~$, and
$(x^m_\sL,\tka,\btka)$ are chiral coordinates of $N{=}(1,1)$ superspace. We
extend the (pseudo)conjugations  \p{ELconj} and \p{sr2conj} to the harmonics by
\be
 \widetilde{u^\pm_k}\=u^{\pm k} \qq\textrm{and}\qq  (u^\pm_k)^*\=u^\pm_k \label{Ooo}
\ee
so that the analytic coordinates conjugate identically for both choices,
\bea
 &&\widetilde{x_\A^m}\=x_\A^m\ ,\qq
\widetilde{\theta^{\pm\alpha}}\=\varepsilon_{\ab}\theta^{\pm\beta}\ ,\qq
\widetilde{\bar\theta^{\pm\da}}\=\varepsilon_{\da\db}\bar\theta^{\pm\da}\ ,
\nn \\[4pt]
 &&(x_\A^m)^*\=x_\A^m\ ,\qq
(\theta^{\pm \alpha})^*\=\varepsilon_{\ab}\theta^{\pm\beta}\ ,\qq
(\bar\theta^{\pm\da})^*\=\varepsilon_{\da\db}\bar\theta^{\pm\da}\
 ;
\eea
in particular, both square to $-1$ on spinor coordinates. This means that
both maps become pseudoconjugations when applied to the extended set
of coordinates. These two
pseudoconjugations act identically on invariants and harmonic superfields,
e.g.~$(A^kB_k)^*=\widetilde{(A^kB_k)}$ or  $(q^+)^*=\widetilde{q^+}$, but differ
on harmonics or R-spinor component fields, e.g.~$(A_k)^*\neq\widetilde{A_k}$.
An important invariant pseudoreal subspace
is the analytic euclidean harmonic superspace, parametrized by the coordinates
\be
\zeta \ \equiv\ (x_\A^m~,~\theta^{+\alpha}~,~\bar\theta^{+\da})~,~u^\pm_i\ .
\ee

The supersymmetry-preserving spinor and harmonic derivatives have
the following form in these coordinates ($\Dp_\sL=\partial^\pp)$:
\bea
&&\Dpa \=\partial^+_\alpha \ ,\qq
\Dma \=-\partial^-_\alpha + 2\ii(\sigma^m)_\ada\btma\partial_m^\A\ ,\nn\\[6pt]
&&\bDpa \=\bar\partial^+_\da \ ,\qq
\bDma \=-\bar\partial^-_\da - 2\ii(\sigma^m)_\ada\tma\partial_m^\A\ ,
\lb{B6b}\\[6pt]
&& \Dp_\A\=\dpp -2 \ii(\sigma^m)_\ada\tpa \btpa \partial_m^\A
+ \tpa \partial^+_\alpha + \btpa\bar\partial^+_\da\ ,
\lb{B6c}
\eea
where $\partial^\pm_\alpha\equiv\partial/\partial\theta^{\mp\alpha}$,
$\bar\partial^\pm_\da\equiv\partial/\partial\bar\theta^{\mp\da}$,
$\partial^\pp = u^{+i}\partial/\partial u^{-i}$.
The partial derivatives in different bases are related as
\bea
&&\partial_m^\sL\=\partial_m^\A\ , \qq
D^{++}_\sL \= \partial^\pp \= D^{++}_\A\ , \nn\\[6pt]
&&\partial^k_\alpha\=-u^{+k}\partial^-_\alpha-u^{-k}\partial^+_\alpha
+2\ii u^{-k}\btpa(\sigma^m)_\ada\partial^\A_m\ , \nn\\[6pt]
&&\bar\partial_{\da k}\=u^+_k\bar\partial^-_\da+u^-_k\bar\partial^+_\da
+2\ii u_k^+\tma(\sigma^m)_\ada\partial^\A_m
\= u^-_k\bDpa-u^+_k\bDma\ .
\lb{pthk}
\eea
A Grassmann analytic superfield is defined by
\be
\Dpa \Phi(\zeta, \theta^-, \bar\theta^-, u) \=
\bDpa \Phi(\zeta, \theta^-, \bar\theta^-, u) \= 0
\ee
and so can be treated as an unconstrained function in the analytic superspace,
$\Phi=\Phi(\zeta,u)$.

It is important to realize that the chirality-preserving operator~$P$
in~\p{Poper} also preserves {\it Grassmann analyticity\/}.
This is seen in the analytic basis using the relations \p{pthk},
\be
\{\partial^k_\alpha, \Dpb \} \= \{\partial^k_\alpha, \bDpb \} \=0
\qq\Longrightarrow\qq [P,\Dpb] \= [P,\bDpb] \= 0 \ .
\ee
In the analytic superspace it takes the form
\be
P \= -\sha  [ \overleftarrow{\partial^-_\alpha}u^{+k}
- 2\ii\overleftarrow{\partial^\A_m}u^{-k}\btpa(\sigma^m)_\ada ]\
C^\ab_{kj}\ [ u^{+j}\overrightarrow{\partial^-_\beta}
- 2\ii u^{-j}\btpb(\sigma^n)_\bdb\overrightarrow{\partial^\A_n}] \ .
\lb{Panal}
\ee
We point out that our deformation is  polynomial (see \p{moyal}),
a property which may also be checked directly in harmonic superspace.
The graded star-commutators in the analytic superspace read
\bea
&&\{\theta^{+\alpha} \stc \theta^{+\beta}\} \=
C^\ab_{kj}u^{+k}u^{+j}\ =: \cC^{\pp\ab}\ ,
\lb{Cdef1} \\[6pt]
&&[x^m_\A \stc \theta^{+\alpha}] \=
2\ii(\sigma^m)_\bdb\btpb C^{\ba}_{kj}u^{-k}u^{+j}\
=:\ \cC^{m+\alpha}\ ,\\[6pt]
&&[x^m_\A \stc x^n_\A] \=
4(\sigma^m)_\ada(\sigma^n)_\bdb\btpa\btpb C^{\ba}_{kj}u^{-k}u^{-j}\
=:\ \cC^{mn}\ . \lb{Cdefanal}
\eea
Note that the functions $\cC^{m+\alpha}$ and $\cC^{mn}$ are nilpotent.
For the singlet deformation \p{Cinv} they simplify to
\bea
&&\cC^{\ab\pp}\=\cC^{mn} \= 0\ ,\qq
\cC^{m+\alpha}\=-2\ii (\sigma^m)^\adb \btp_\db\,I \ ,\lb{Idef}\\[6pt]
&&\Longrightarrow\qq
P_{singlet}\=-\ii (\sigma^m)^\adb \btp_\db\,I\,(
\overleftarrow{\partial^\A_m}\, \overrightarrow{\partial^-_\alpha} -
\overleftarrow{\partial^-_\alpha}\, \overrightarrow{\partial^\A_m} )
\eea
which satisfies $P_{singlet}^3=0$.
Since $P$ is independent of harmonics, $[D^{\pm\pm},P]=0$,
the deformed coordinate (anti)commutators are closed under the action of
$D^{\pm\pm}$.
The transformation properties of analytic superfields are standard
like in the case of chiral superfields. As in the previous section,
the deformation operator $P$ generically breaks $N{=}(1,1)\to N{=}(1,0)$
supersymmetry and part of the euclidean and R~symmetries.
Note, however, that on analytic superfields the degenerate deformation
$P_{deg}(Q^2)$ contains only
\be
Q^2_\alpha \= -u^{+2}\partial^-_{\alpha}
+2\ii u^{-2}\btpa (\sigma^m)_{\alpha\da}\partial_m^\A\ .
\ee
Hence, this deformation preserves chirality, analyticity and $N{=}(1,\ha)$
supersymmetry.

If we do not care about chirality we may add to $P$ any one of
the two supersymmetry-preserving operators \p{LRdef} which in analytic
coordinates read
\be
L \= \sha \bigl(
\olD{}^{+\alpha} J \orD{}^{-}_\alpha + \olD{}^-_\alpha  J \orD{}^{+\alpha}
\bigr)\ , \qq
R \= \sha \bigl(
\olbD{}^{+\da} \bar{J} \orbD{}^-_\da + \olbD{}^-_\da \bar{J} \orbD{}^{+\da}
\bigr)\ .
\lb{Lanal}
\ee
These operators do not deform products of analytic superfields
$\Phi(\zeta, u)$ and $\Lambda(\zeta, u)$:
\be
\Phi\,\e^L \Lambda \= \Phi\,\Lambda\ , \qq
\Phi\,\e^R \Lambda \= \Phi\,\Lambda\ .
\ee

Our choice of deformation \p{Poper} was a minimal one for chiral superspace
coordinates, in the sense that we took $x^m_\sL$ and $\btka$ to
star-(anti)commute with anything. Alternatively, one can conceive of
minimal deformations for analytic superspace coordinates.
For such {\it analytic nilpotent deformations\/},
the graded (anti)commutators of all analytic superspace coordinates
$\left( x^m_\A, \theta^{+\alpha}, \bar\theta^{+\dot\alpha}, u^{\pm i} \right)$
among themselves vanish, except perhaps for
\be
\{ \theta^{+\alpha} \stc \theta^{+\beta} \} \=
\hat{\cC}^{++\ab} \ =:\ \hat{C}^{\ab}_{ik}u^{+i}u^{+k}\ ,
\lb{basic2}
\ee
where $\hat{C}^{\ab}_{ik}$ are constants.
This does not yet fix the deformation since nothing was said about
$\theta^{-\alpha}$ or $\bar\theta^{-\dot\alpha}$.
It is the natural assumption that
\be
D^{\pm\pm}_\sL x^m_\sL\ \equiv\ \partial^{\pm\pm}x^m_\sL \= 0
\qq\textrm{and}\qq
\partial^{\pm\pm} \theta^\alpha_k \=
\partial^{\pm\pm}\bar\theta^{\dot\alpha}_k \= 0
\lb{harmindep1}
\ee
which uniquely determines the analytic nilpotent deformation.
First, it maps~\p{basic2} to
\be
\{ \theta^{+\alpha} \stc \theta^{-\beta} \} \=
\hat{C}^{\ab}_{ik}u^{+i}u^{-k}\ .\lb{basic3}
\ee
Second, requiring
\be
[x^m_\A \stc x^n_\A] \= 0 \qq \textrm{and} \qq
[x^m_\A \stc \theta^{+\alpha}] \= 0\ ,
\lb{ana1}
\ee
relating $x^m_\A$ to $x^m_\sL$ via \p{ancoor}
and using \p{basic3}, the harmonic-independence assumption~\p{harmindep1}
straightforwardly yields $\hat{\cC}^{++\ab}=0$, thus reducing
$\hat{C}$ to its SO$(4)\times\,$SU(2) singlet part,
$\hat{C}^{\ab}_{ik} = \varepsilon^{\ab}\varepsilon_{ik} J$.
Therefore, we arrive at
\be
\{\theta^{+\alpha} \stc \theta^{+\beta}\} \= 0 \qq\textrm{and}\qq
\{\theta^{+\alpha} \stc \theta^{-\beta}\} \= \varepsilon^{\ab}\,J
\lb{ana2}
\ee
for this case.
Third, the remaining graded (anti)commutators are easily reconstructed to be
\be
\{\theta^{-\alpha} \stc \theta^{-\beta}\} \= 0 \qq\textrm{and}\qq
[ x^m_\A \stc \theta^-_\beta ] \=
-2\ii(\sigma^m)_{\beta\dot\beta} \bar\theta^{-\dot\beta}\,J\ .
\lb{ana3}
\ee
Comparison with \p{Lanal} shows that \p{ana1}--\p{ana3} are precisely
and uniquely generated by the supersymmetry-preserving deformation
operator $L$. To summarize, nilpotent analytic deformations correspond to~$L$.
At the same time, chirality is no longer preserved, since e.g.
\be
[x^m_\sL \stc \theta_{\beta i}] \=
-2\ii (\sigma^m)_{\beta\dot\beta} \bar\theta^{\dot\beta}_i\,J
\ee
and $x^m_\sL$ ceases to commute with itself as well.

Apparently, any theory which admits a formulation entirely in analytic harmonic
superspace will stay undeformed under an analytic nilpotent deformation as
defined above. This statement applies in particular to hypermultiplets.
In contrast, theories which require the full $N{=}2$ superspace for their
off-shell formulation, e.g.~gauge theories, will experience specific
deformations.

Finally, we note that it is possible to define more general deformations
in the analytic basis of $N{=}2$ superspace if one denies the
harmonic-independence conditions~\p{harmindep1}. Although such deformations
are worth to be investigated, the deformations respecting $[D^{\pm\pm},P]=0$
are distinguished in that they retain a link with the standard $N{=}2$
superspace. Note also that we do not analyze here deformations based on the
SU(2)~breaking Grassmann-analytic coordinates
$\theta^{1\alpha}{+}\ii\theta^{2\alpha}$ which were considered in \cite{FLM}.

\setcounter{equation}0
\section{Interactions in deformed harmonic superspace}

Harmonic superspace with noncommutative bosonic coordinates $x^m_\A$ has been
discussed in~\cite{BS}. This deformation yields nonlocal theories
but preserves the whole $N{=}2$ supersymmetry. In contrast, we expect that
the deformations defined in the previous section will produce much weaker
nonlocalities due to their nilpotency. Leaving quantum considerations for
future study, we present here the chirally deformed versions of the
off-shell actions for some basic theories in harmonic superspace.

We shall limit our attention to the deformation operator $P$ which affects
analytic superfields and preserves both analyticity and chiralities.
The free $q^+$ and $\omega$ hypermultiplet actions of ordinary harmonic theory
\cite{GIOS} are not deformed in non\-(anti)\-commutative superspace:
\be
S_0(q^+)\=\int\!\diff u\,\diff\zeta^{-4}\ \tilde{q}{\,}^+ \Dp q^+\ ,\qq
S_0(\omega)\=\int\!\diff u\,\diff\zeta^{-4}\,(\Dp\omega)^2 \lb{freeAc}\ ,
\ee
where $\diff\zeta=\diff^4x_\A (D^-)^4$.
Non\-(anti)\-commutativity arises in interactions, for instance
for the self-interaction of the hypermultiplet,
\be
S_4(q^+)\=\int\!\diff u\,\diff\zeta^{-4}\,
(a\ \tilde{q}{}^+ \star q^+ \star \tilde{q}{}^+ \star q^+\ +\
 b\ q^+ \star q^+ \star \tilde{q}{}^+ \star \tilde{q}{}^+ )\ ,
\ee
where $a$ and $b$ are real coupling constants.
Expanding out the star products yields a finite number of corrections
to the local interaction term $(q^+\tilde{q}{}^+)^2$.
As an example, for the simplest nilpotent deformation~\p{Idef} one has
\be
\tilde{q}{}^+ \star q^+ \=
\tilde{q}{}^+ (1+P_{singlet}+\sha P_{singlet}^2)\,q^+\ .
\ee
An mentioned before, these corrections explicitly break a part of the
symmetries~\p{defQ}.

The interaction of the hypermultiplet~$q^+$ with a U(1) analytic gauge
superfield~$\Vp$ can be introduced as in \cite{BS}, by replacing $\Dp$
in~\p{freeAc} with the covariant harmonic noncommutative left-derivative,
\be
\Dp q^+ \qq \Longrightarrow \qq \nabla^\pp q^+\=\Dp q^+ + \Vp \star q^+\ .
\lb{defD}
\ee
The gauge transformation of the anti-Hermitian $\Vp$ reads
\be
\delta_\lambda\Vp\=-\Dp \lambda+[\lambda \stc \Vp]
\ee
where $\lambda$ is an anti-Hermitian analytic gauge parameter.
The generalization to U($n$) analytic gauge fields is straightforward.
Note again that from the beginning we retain only those symmetries
which are unbroken by the deformation of choice.

In Wess-Zumino gauge we have
\bea
\Vp_{\W\Z} &=& (\tp)^2\bar\phi\ +\ (\btp)^2\phi\ +\
\tpa\btpa A_\ada \nn\\[4pt]
&+& (\tp)^2\btpa u^-_k \bar\lambda^k_\da\ +\
(\btp)^2\tpa u^-_k\lambda^k_\alpha\ +\ (\tp)^2(\btp)^2u^-_ku^-_jX^{kj}\ ,
\lb{WZ}
\eea
with all components being functions of~$x^m_\A$,
and a component expansion of the hypermultiplet~$q^+$ which consists
of infinitely many terms due to the harmonic dependence.
The component expansion of the deformed products is rather complicated
since the number of terms increases significantly.
For the singlet deformation~\p{Idef}, the star product in~\p{defD}
contains the terms
\bea
\Vp P_{singlet}\,q^+\!&=&\!\sha\cC^{m\alpha+}
[\partial_m^\A\Vp\partial^-_\alpha q^+-\partial^-_\alpha\Vp\partial_m^\A q^+ ]
\ ,\\[6pt]
\Vp P_{singlet}^2\,q^+\!&=&\!-\sfrac14\cC^{m\alpha+}\cC^{n\beta+} \\[4pt]
&&\times [\partial_m^\A\partial_n^\A\Vp\partial^-_\alpha \partial^-_\beta q^+
+\partial^-_\alpha \partial^-_\beta\Vp\partial_m^\A\partial_n^\A q^+
+\,2\partial^-_\beta \partial^\A_m\Vp\partial_n^\A\partial_\alpha^- q^+]\ .\nn
\eea
Since $\cC^{m+\alpha}=-2\ii(\sigma^m)^\adb\btp_\db I$,
all terms proportional to $(\btp)^2$ in the component expansion of
$\Vp$ or $q^+$ drop out of this deformation. In particular, the
auxiliary field $X_{kj}$ of the vector multiplet~\p{WZ} does not appear.
On the other hand, space-time derivatives of propagating fields do occur;
for instance, the term $\Vp P_{singlet}\,q^+$ produces
\be
\ii I(\sigma^m)_\adb\btpb(\tp)^2\,\partial_m^\A\bar\phi\,\psi^\alpha~.
\ee

The action for this noncommutative U(1) gauge superfield can be constructed
in central coordinates in analogy with the action for commutative $N{=}2$
Yang-Mills theory~\cite{Z1}, but it is easier to analyze it in chiral
coordinates.
Following~\cite{Z1}, one constructs the deformed connection for the
derivative $\Dm$ via
\bea
&&\Dp\Vm-\Dm\Vp+[\Vp \stc \Vm]\=0\ ,\\[6pt]
&&\Vm(z_\sL, u)\=
\sum\limits_{n=1}^\infty(-1)^n\!\int\!\diff u_1\ldots\diff u_n
\frac{\Vp(z_\sL, u_1) \star \Vp(z_\sL, u_2) \ldots \star \Vp(z_\sL, u_n)}
{(u^+u^+_1)(u^+_1u^+_2)\ldots (u^+_nu^+)}\ ,\nn
\eea
where $(u^+_1u^+_2)^{-1}$ is a harmonic distribution (see \cite{GIOS2,GIOS}).
In general, the action for~$\Vp$ contains an infinite number of vertices,
with star commutators substituting the ordinary commutators of~$\Vp$
taken from the standard non-abelian action.
On the cubic level, for example, one obtains
\be
S_3(\Vp)\=\frac{1}{3g^2}\int\!
\diff u_1\,\diff u_2\,\diff u_3\,\diff^4x_\sL\,\diff^8\theta\,
\frac{\Vp(z_\sL, u_1) \star \Vp(z_\sL, u_2) \star \Vp(z_\sL, u_2)}
{(u^+_1u^+_2)(u^+_2u^+_3)(u^+_3u^+_1)}\ ,\lb{3int}
\ee
where $g$ is the gauge coupling.
The chiral and anti-chiral superfield strengths $\cW$ and $\bcW$
in the euclidean case are independent. They have the form
\be
\cW\= -\, \sfrac14(\bar D^+)^2\Vm\ , \; \bcW= -\, \sfrac14(D^+)^2\Vm\ ,\;
\textrm{with}\;\;\delta_\lambda\,(\cW, \bcW)\=[\lambda \stc (\cW, \bcW)]\ ,
\ee
and satisfy the covariantized chirality and harmonic-independence conditions
\be
\bDpa\cW\ = 0\ ,\q
\bDma\cW-[\bDpa\Vm \stc \cW]\=0\ ,\q
\Dp\cW + [\Vp \stc \cW] \=0\ ,
\lb{covharm}
\ee
plus analogous conditions on the anti-chiral $\bcW$,
as well as $\ (D^+)^2\cW = (\bar D^+)^2\bcW$.
For the case of the chirality-preserving deformations, one can write down
gauge-invariant actions holomorphic in~$\cW$, such as
\be
S_{\cW}\ \sim\ \int\!\diff^{4}x_\sL\,\diff^4\theta\
\cW \star \cW \star\cW\ , \lb{Abel1}
\ee
and likewise for the covariant anti-chiral superfield strength.
It is easy to check that
\be
\delta_\lambda\, S_{\cW} \= 0 \qq\textrm{and}\qq
\Dp\, S_{\cW} \= \bar D_{\da k}\,S_{\cW} \= 0\ .
\ee

In the Feynman rules, the only effect of our deformations is a small number
of higher-derivative contributions to the standard interaction vertices.
Due to the nilpotency of these deformations, the locality of the theory
is not jeopardized. It should be straightforward to evaluate the ensueing
mild corrections to the known quantum properties of $N{=}2$ harmonic superspace.

\setcounter{equation}0
\section{Conclusions}
We have considered nilpotent deformations of $N{=}(1,1)$ chiral and
Grassmann-analytic harmonic superfields. A minimal setup deforms only
the fermionic coordinates and keeps the bosonic ones entirely commuting.
Applied to chiral superspace coordinates, we call such a deformation
{\it chiral nilpotent\/} because it preserves chirality.
On the background of non-deformed euclidean $N{=}(1,1)$ superspace, one can
treat it as a soft breaking of the part of supersymmetry and automorphism
symmetry which is generated by $\bar Q_{\da k}$, $L^\alpha_\beta$ and $L^k_j$,
respectively.
Up to a choice of basis, the constant deformation matrix is determined by
four parameters. Interestingly, a special choice of these retains the
SO(4)$\,\times\,$SU(2) automorphism invariance but still breaks
$N{=}(1,1)\to N{=}(1,0)$ supersymmetry. Yet, contrary to an assertion
of a recent preprint~\cite{last}, it is possible to keep a fraction of
$\frac34$, i.e.~$N{=}(1,\ha)$ supersymmetry, by employing a {\it degenerate\/}
deformation matrix. Complete supersymmetry, however, can only be saved at
the expense of chirality.

As the main new development, we extended the analysis to euclidean $N{=}(1,1)$
{\it harmonic\/} superspace,
parametrized by analytic coordinates. For these, chiral nilpotent deformations
induce nilpotent noncommutativity also in the bosonic sector but preserve
Grassmann analyticity. It is therefore consistent to deform only the analytic
subspace in this manner. As an alternative, we also investigated the minimal
situation in analytic coordinates. Such {\it analytic nilpotent\/}
deformations may respect full supersymmetry but violate chirality.
It turned out that they do not affect the analytic subspace but only the
central superspace; hence, they leave hypermultiplets undeformed.
Finally, we gave examples of superfield theories in chiral-nilpotently
deformed harmonic superspace. In particular, we have shown how to construct
the SO(4)$\,\times\,$SU(2) invariant nilpotent deformation of $N{=}(1,1)$
supersymmetric U(1) gauge theory in chiral superspace coordinates.

It would be interesting to understand a possible string theoretic origin
of the deformations considered here.
On the more technical side, one may work out the effect of these deformations
on the component actions and perform quantum calculations, e.g.~for obtaining
effective actions, in the off-shell harmonic superspace approach.
\vspace{0.3cm}

\noindent{\bf Note added.} Soon after the first version of our paper was listed
on hep-th, there appeared an e-print~\cite{FS} also addressing nilpotent
singlet deformations of $N{=}2$ harmonic superspace and construction of the
corresponding field models.

\section*{Acknowledgments}
This work was partially supported by the DFG priority program SPP 1096 in
string theory, INTAS grant No 00-00254, RFBR-DFG grant No 02-02-04002,
grant DFG No 436 RUS 113/669, RFBR grant No 03-02-17440 and a grant of the
Heisenberg-Landau program.

\vspace{0.5cm}


\end{document}